\documentclass{elsart}

\usepackage{amssymb}
\usepackage{graphicx}

\begin{document}

\begin{frontmatter}

\title{PROCRUSTES: A computer algebra package for post-Newtonian calculations in General Relativity}

\author{Dirk Puetzfeld}
\ead{dirk.puetzfeld@astro.uio.no}
\ead[url]{www.thp.uni-koeln.de/$\sim$dp}
\address{Institute of Theoretical Astrophysics, University of Oslo, P.O.\ Box 1029, 0315 Oslo, Norway}

\begin{abstract}
We report on a package of routines for the computer algebra system Maple which supports the explicit determination of the geometric quantities, field equations, equations of motion, and conserved quantities of General Relativity in the post-Newtonian approximation. The package structure is modular and allows for an easy modification by the user. The set of routines can be used to verify hand calculations or to generate the input for further numerical investigations. 
\end{abstract}

\end{frontmatter}


\begin{small}
\noindent
{\bf Program summary}
\label{program_summary_sec}

\textit{Title of the program:} Procrustes \newline

\textit{Catalogue identifier:} ADYH\_v1\_0 \newline

\textit{Program obtainable from:} CPC Program Library, author's webpage\newline

\textit{Computer for which the program is designed and others on which it has been tested:}\newline
\textit{Computers:}\newline
 Platforms supported by the Maple computer algebra system (program was written
under Maple 8, but also tested with Maple 9, 9.5, 10) \newline 

\textit{Operating systems under which the program has been tested:} \newline 
Linux, Unix, Windows XP\newline

\textit{Programming language used:}\newline
Maple internal language \newline

\textit{Memory required to execute typical problem:}\newline
Dependent on problem (small $\sim$ couple of MBytes, large $\sim$ several GBytes)\newline

\textit{Classification:}\newline
1.5  Relativity and Gravitation, 5 Computer Algebra

\textit{No.\ bits in a word:}\newline 
Dependent on Maple distribution (supports 32 bit and 64 bit platforms)\newline

\textit{No.\ of processors used:} 1\newline

\textit{No.\ of bytes in distributed program, including test data, etc.:} $\sim$150 kbyte\newline

\textit{Distribution format:} Compressed tar file, compressed zip archive\newline

\textit{Nature of the physical problem:} \newline
The post-Newtonian approximation represents an approximative scheme frequently used in General Relativity in which the gravitational potential is expanded into a series in inverse powers of the speed of light. Depending on the desired approximation level the field equations and equations of motion have to be determined up to given orders in the speed of light. This usually requires large algebraic computations due to the geometrical quantities entering the field equations and equations of motion.\newline 

\textit{Method of solution:}\newline  
Automated computation using computer algebra techniques. Program has modular structure and only makes use of basic features of Maple to guarantee maximum compatibility and to allow for rapid extensions/modifications by the user.\newline

\textit{Typical running time:} \newline
Dependent on problem (small $\sim$ couple of minutes, large $\sim$ couple of hours)\newline

\textit{Restrictions on the complexity of the problem:}\newline
Sufficient amount of memory is the limiting factor for these calculations. The structure of the program allows one to handle large scale problems in an iterative manner to minimize the amount of memory required.
\end{small}

\section{Introduction}
\label{Introduction_sec}

Due to the structure of the field equations and the geometrical quantities entering these equations, computations in General Relativity are usually very tedious and prone to errors if performed by hand. This is especially true for calculations in which one makes use of some approximation scheme, such as series expansions. The so-called post-Newtonian approximation, in which certain quantities are developed in powers of the speed of light, is an example of such an expansion scheme. The post-Newtonian approximation was used in many works to determine the equations of motion in an iterative way \cite{LorentzDroste:1917,Fock:1959,Chandrasekhar:1965:2}, see also \cite{Puetzfeld:2006} for a more comprehensive list, and plays an important role in the experimental verification of General Relativity \cite{Will:1993}.

In contrast to the efforts in the field of exact solutions, in which computer algebra methods are heavily used, and several specialized packages \cite{Mathtensor:link,Eins:link,Excalc:link,Cartan:link,Grtensor:link} for common computer algebra systems \cite{Maple:link,Mathematica:link,Reduce:link,Mupad:link} exist, there do not seem to be many works which rely on the use of such methods when working in the post-Newtonian approximation. One example of a package which has been utilized in a post-Newtonian context can be found in \cite{Klioner:1998}. Further literature on the use of computer algebra methods in General Relativity as well as a list of special purpose systems is given in \cite{HehlPuntigamRuder:1996,Grabmeier:2003}.

In this work we present a set of routines for the computer algebra system Maple \cite{Maple:link}. The routines can be used to calculate the explicit form of the field equations, equations of motion, and conserved quantities in an automated fashion. The order of approximation is controlled by a set of switches and allows for a quick modification by the user. The output produced by the program can be used to support and verify hand calculations or to provide the input for subsequent numerical analysis. A splitting of time and resource intensive calculations into several small subproblems is supported by the modular structure of the routines. 

The structure of the present work is as follows. In section \ref{structure_sec} we provide a general overview of the package and a listing of different types of routines. Subsequently, we discuss a typical application in section \ref{application_sec}, where in we also give several explicit examples on the usage of different routines in the package and discuss their input/output. In section \ref{conclusions_sec} we summarize the details of our package and discuss possible future extensions and modifications. Appendix \ref{functions_sec} contains a comprehensive directory of switches and functions available in Procrustes.

\section{Program structure}
\label{structure_sec}

\subsection{General thoughts}
\label{thoughts_sec}

The set of routines which we summarize here under the name {\it Procrustes} emerged from the author's need to tackle a specific computation \cite{Puetzfeld:2006} within the post-Newtonian approximation. They were not designed to cover every imaginable problem in such a realm. The original task was to explicitly calculate the geometrical objects, field equations, and equations of motion for a given metric ansatz and to display the results in the form of a series expansion. 

Since our goal was to create a program which is easy to understand and modify, we did {\it not} make use of any of the existing systems for tensor manipulations. Consequently, there is {\it no} built-in differentiation between upper and lower indices of objects, just operations with lists.
 
Because of the ease of use of algebraic manipulations and simplifications, we chose to implement our routines in the Maple \cite{Maple:link} system. This program is available for many platforms and operating systems, supports large amounts of memory, has a large set of built in routines for symbolic computations/ simplifications, and supports all kinds of import and export functions. We only make use of very basic functions of Maple, especially its flexible list data structure, and intentionally use only the simplest save command, instead of the generation of a package repository which leads to a version dependence, to store our set of routines. 

\subsection{Routine and program structure}
\label{routine_sec}

As mentioned in the introduction we are concerned with the post-Newtonian approximation of General Relativity. Since General Relativity represents a metric theory of gravitation, subsequent geometrical objects can be deduced from the metric. In the case of the post-Newtonian approximation the metric is developed into a series of inverse powers of the speed of light $c$ starting from the Newtonian limit, i.e.\ formally we have $g_{\alpha \beta }=g_{\alpha \beta }^{\rm {\tiny \tiny Newton}}+c^{-1}\stackrel{1}{g_{\alpha \beta}}+c^{-2}\stackrel{2}{g_{\alpha \beta}}+\dots$ The strategy within the post-Newtonian approximation then typically consists of rewriting the field equations in a form which closely resembles the structures which we already know from Newtonian gravity. In addition the slow-motion condition employed in the post-Newtonian approximation leads to a hierarchy between the different metric components when one considers a split into temporal and spatial components. The assumption of slow motions also motivates the series expansion of the velocities, which enter the energy-momentum tensor, by inverse powers of the speed of light, i.e.\ $u^{\alpha}= \stackrel{0}{u^{\alpha}} + c^{-1}\stackrel{1}{u^{\alpha}} + c^{-2}\stackrel{2}{u^{\alpha}}+\dots$ We are not going into further details of the post-Newtonian approximation here but refer the reader to the overviews \cite{Fock:1959,InfeldPlebanski:1960,Damour:1987,Soffel:1989,Brumberg:1991} and to the extensive literature list given in \cite{Puetzfeld:2006}. 

Our main concern, in this article, is the introduction of expansions for the metric $g_{\alpha \beta}$ and velocity $u^\alpha$, in inverse powers of the speed of light, in all subsequent quantities to be calculated. In addition we must take care of the fact that derivatives with respect to the time coordinate increase the order of smallness of a term by $c$. 

\begin{figure}
\begin{center}
\includegraphics[width=10cm]{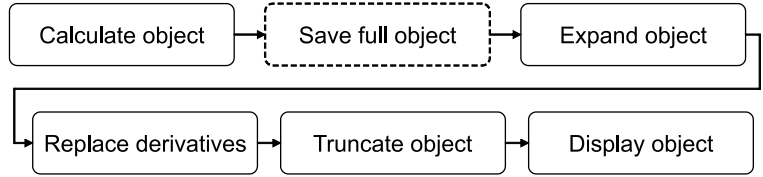}
\end{center}
\caption{Sketch of a routine which calculates and displays an object. The dashed line indicates that saving the object to a file is optional.} 
\label{calc_scheme_fig}
\end{figure}

We handle the increase of the order of smallness via derivative replacements. In all routines the replacement is performed after the full object has been calculated and only plays a role in the expression which is subsequently used to produce the output (see also figure \ref{calc_scheme_fig}). The display order of the output is directly passed to the routine \texttt{calc\_<object name>(<display order>)}. Only the full object remains in memory. Depending on the definition of the metric and other matter variables, which enter through energy-momentum tensor of the system under consideration, the user has to supply a list of substitution rules for all occurring derivatives with respect to the time coordinate for these variables. This substitution list has to be provided in the global variable \texttt{sublist}. We note that for the correct operation of the substitution procedure it is mandatory that \texttt{sublist} contains a hierarchical list of time derivative replacements starting with the highest expected derivative. In order to help the user with the generation of such a substitution list the package contains a small routine which produces a valid list for a given set of variable names (see the corresponding section in the sample worksheet provided with the package).  

\begin{figure}
\begin{center}
\includegraphics[width=10cm]{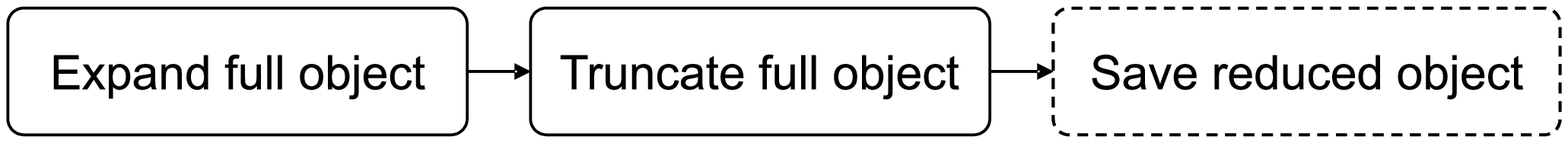}
\end{center}
\caption{Sketch of a routine which reduces an object. The dashed line indicates that saving the object to a file is optional.} 
\label{red_scheme_fig}
\end{figure}

In addition to the routine which performs the actual calculation of an object, named \texttt{calc\_<object name>}, there is also a routine, named \texttt{red\_<object name>}, which performs a truncation of the object at the specified order. The truncation routines should be used in order to reduce the amount of memory needed in the calculation of subsequent objects, such as the reduction of the connection before the curvature is calculated. The general structure of the reduction routines is sketched in figure \ref{red_scheme_fig}. The display order of the output is directly passed to the routine \texttt{red\_<object name>(<display order>)}. No derivative substitution is performed and only the truncated version of the full object remains in memory. Note that the reduction routines do {\it not} make use of the substitution list. 

The routines in the Procrustes package are simple procedures which operate on a set of global variables; see tables \ref{function_list_table_0}--\ref{function_list_table_4} for a list of routines and table \ref{global_variables} for a list of variables. All routines are stored and loaded from a subdirectory defined in the global variable \texttt{CCprocrustes\_dir}. We intentionally used Maple's \texttt{save} command to store the routines, instead of creating a package repository which leads to a dependence on the underlying Maple version. All of the routines can be found in the file \texttt{procrustes\_<version number>.mws}, the user may modify this file to replace existing or add his/her own routines. A simple execution of this worksheet will store the routines in the package subdirectory.

\begin{figure}
\begin{center}
\includegraphics[width=10cm]{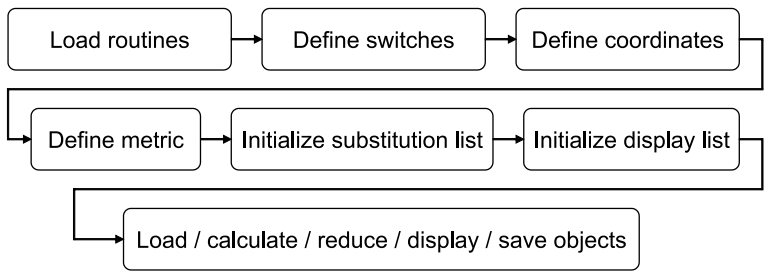}
\end{center}
\caption{Preferred structure for a calculation.}
\label{prog_scheme_fig}
\end{figure}

The recommendation for the general structure of a worksheet to be used with the package is sketched in figure \ref{prog_scheme_fig}. In the next section we provide some detailed comments on the input and output of the demonstration worksheet \texttt{procrustes\_demo.mws} supplied together with the package. 

\section{Typical application}
\label{application_sec}

In this section we comment on a sample calculation with Procrustes. The input/output displayed here can be produced with the worksheet \texttt{procrustes \_demo.mws} supplied with the package. We note that this worksheet allows one to quickly reproduce many of the results of the classic paper \cite{Chandrasekhar:1965:2}.

\subsection{Installation}

To install the package create a directory of your choice and, depending on the operating system, unpack the \texttt{tar} or \texttt{zip} archive. The directory should contain two files \texttt{procrustes\_<version number>.mws} (containing the source of the package), and \texttt{procrustes\_demo.mws} (a demonstration worksheet; excerpts of this file are discussed in the sections below). The subdirectories \texttt{procrustes<version number>} (containing the set of routines generated from the source file), and \texttt{testrun\_procrustes<version number>} (containing data from a testrun of the package, which is used by the worksheet to determine if the package is installed correctly).

Modifications of the package can be performed by editing the source of the routines in \texttt{procrustes\_<version number>.mws}. When introducing new routines we recommend maintaining the same naming convention as introduced in the original file (see also the complete list of routines in tables \ref{function_list_table_0}-\ref{function_list_table_4}). New routines should be added to the save list at the end of the file in order to guarantee their storage for further usage. A simple execution of the worksheet then stores the modified/new routines in the package subdirectory. 

In an actual calculation we recommend to load the required set of routines from the package subdirectory. The worksheet \texttt{procrustes\_demo.mws} contains, apart from some instructive sections on how to specify metrics and substitution lists, a sample calculation which sticks to the scheme sketched in figure \ref{prog_scheme_fig}.   

\subsection{Coordinate and metric specification}

The coordinates to be used by Procrustes have to be specified in form of a list in a global variable named \texttt{xx}, in the first section of the worksheet this is done by the command:\footnote{In the following input by the user is marked by ``\texttt{>}'' and output by Maple is preceeded by ``\texttt{<}''. Three vertically aligned dots are used to indicated that we shortened the displayed input/output.}
\begin{tiny}
\begin{verbatim} 
> xx:=[x0,x1,x2,x3];
\end{verbatim}
\end{tiny}
Based on these coordinates the metric has to be specified in form of a list. The worksheet contains a section in which the contravariant as well as the covariant form of the metric are specified, the latter one \texttt{gdd} being defined as:  
\begin{tiny}
\begin{verbatim}
> gdd:=[[1-2*U(xx[1],xx[2],xx[3],xx[4])/c^2+phi(xx[1],xx[2],xx[3],xx[4])/c^4,hdd01(xx[1],xx[2],xx[3],xx[4]/c^3,
         hdd02(xx[1],xx[2],xx[3],xx[4]/c^3,hdd03(xx[1],xx[2],xx[3],xx[4])/c^3],
        [hdd01(xx[1],xx[2],xx[3],xx[4])/c^3,(-1-2*V(xx[1],xx[2],xx[3],xx[4])/c^2),0,0],
        [hdd02(xx[1],xx[2],xx[3],xx[4])/c^3,0,(-1-2*V(xx[1],xx[2],xx[3],xx[4])/c^2),0],
        [hdd03(xx[1],xx[2],xx[3],xx[4])/c^3,0,0,(-1-2*V(xx[1],xx[2],xx[3],xx[4])/c^2)]];
\end{verbatim}
\end{tiny}
This definition corresponds to the line element 
\begin{eqnarray}
ds^2&=&g_{\alpha \beta} dx^\alpha dx^\beta \nonumber \\
    &=&\left(1-\frac{2U}{c^2}+\frac{\phi}{c^4}\right) dx^0 dx^0+\frac{2h_{0a}}{c^3} dx^0 dx^a-\left( 1+\frac{2V}{c^2} \right) \delta_{ab} dx^a dx^b,
\end{eqnarray}
where $U$, $h_{0a}$, and $V$ represent free functions which depend on the coordinates indicated. We remind the reader that there is no built-in distinction between upper and lower indices, hence also the contravariant form \texttt{guu} of the metric has to be provided by hand (see sample worksheet). Coordinate as well as metric definitions are then saved to the files \texttt{coordinates\_demo.mpi, gdd\_demo.mpi, guu\_demo.mpi} for later usage in the sample calculation.

\subsection{Substitution and display list specification}

As we already have mentioned in section \ref{structure_sec} the increase in the order of smallness of quantities is handled via derivative replacements. The substitution list has to be provided in a global variable named \texttt{sublist} and should contain a hierarchical list of replacement rules, starting with highest expected derivative of a variable. In the sample worksheet we have included a section in which a substitution list  is explicitly specified. This list is subsequently used in the sample calculation and has the following structure:    
\begin{tiny}
\begin{verbatim} 
> sublist:=[diff(U(x0,x1,x2,x3),x0,x0,x0,x0)=diff(U(t,x1,x2,x3),t,t,t,t)/c^4,
            diff(U(x0,x1,x2,x3),x0,x0,x0)=diff(U(t,x1,x2,x3),t,t,t)/c^3,
            diff(U(x0,x1,x2,x3),x0,x0)=diff(U(t,x1,x2,x3),t,t)/c^2,
            diff(U(x0,x1,x2,x3),x0)=diff(U(t,x1,x2,x3),t)/c,
            U(x0,x1,x2,x3)=U(t,x1,x2,x3), 
            .
            .
            .
\end{verbatim}
\end{tiny}
We have added a procedure called \texttt{generate\_sublist} to assists the user in the generation of a correct substitution list for all variables in a calculation.

Apart from the substitution list, the package also makes use of a display list, to be defined in the global variable \texttt{displist}. The purpose of this list is to bring the output from the routines in the package into a neat form which can be quickly interpreted by the user. The display list does {\it not} play a role in the actual calculation of quantities, nor in the storage of partial results, it is only used when a procedure outputs an object on the screen (see figure \ref{calc_scheme_fig}). In the sample worksheet provided the display list is used to suppress the functional dependencies of the variables in the calculation, thereby reducing the length of the output considerably. The example for \texttt{displist} given therein has the following structure:
\begin{tiny}
\begin{verbatim} 
> displist:=[U(t,x1,x2,x3)=U,
             V(t,x1,x2,x3)=V,
             rho(t,x1,x2,x3)=rho,
             pp(t,x1,x2,x3)=p,
             .
             .
             . 
\end{verbatim}
\end{tiny}
Depending on the needs of the user one could also use the display list to pre-process output which can then be directly exported into \LaTeX. Either by exclusively using the display list or by a combination of the display list and Maple's built-in output facility. In the next section we show some input and output examples from the worksheet. We caution the user that quantities which were reduced by the substitutions in the display list should {\it not} be used for any further calculations.  

\subsection{Calculation and reduction of objects}

After the specification of the coordinates, metric, substitution list, and display list we are ready to perform a calculation with the package. In the demonstration worksheet we provide a sample for the general structure of a worksheet for calculations with Procrustes. After a preamble in which we load the desired procedures from the package directory, and the specification of some global switches (see the list in table \ref{global_switches}), we make use of the previous definitions of the metric and the corresponding substitution list by loading them from the previously specified files and then perform some actual calculations. For example, the command
\begin{tiny}
\begin{verbatim}
> calc_uu(4);
\end{verbatim}
\end{tiny}
calculates and displays the contravariant four velocity up to third order in inverse powers of the speed of light $c$,
\begin{tiny}
\begin{verbatim}
<                            2           2       2
                         vu03        vu01    vu02
                         ----- + U + ----- + -----
                           2           2       2
 Velocity up[, 0, ], 1 + -------------------------
                                      2
                                     c


                                3            2                     2
                            vu01    vu01 vu03             vu01 vu02
                            ----- + ---------- + vu01 U + ----------
                     vu01     2         2                     2
 Velocity up[, 1, ], ---- + ----------------------------------------
                      c                         3
                                               c
  .
  .
  .
\end{verbatim}
\end{tiny}
This corresponds to the result
\begin{eqnarray}
u^0 = 1+\frac{1}{c^2} \left( \frac{1}{2}\delta_{ab}v^a v^b+U \right), \quad u^1 = \left[ \frac{1}{c}+\frac{\frac{1}{2}\delta_{ab}v^a v^b+U}{c^3} \right] v^1, \quad \dots
\end{eqnarray}
In addition to the actual calculation of objects it is sometimes advantageous to reduce the order of an object, i.e.\ to remove higher orders in its series expansion which are not needed in subsequent calculations. Such a reduction is supported by functions starting with \texttt{red\_<object name>}, which exist for most of the objects calculable with the package, cf.\ table \ref{function_list_table_3}. In contrast to the routines which calculate objects and display them up to a given order, the reduction routines truncate the expanded version of the full object before derivative substitutions are performed, see also the scheme sketched in figure \ref{red_scheme_fig}. As an example we calculate the connection from the sample metric,  
\begin{tiny}
\begin{verbatim}
> calc_condndndn(4);

<                                                   d
                                                    -- U
                                                    dt
             Connection (condndndn)[, 0, 0, 0, ], - ----
                                                      3
                                                     c
 .
 .
 .
\end{verbatim}
\end{tiny}
Note that the order passed to the calculation command only influences up to the order objects are displayed, {\it not} the actual form of the object which is kept in memory for further calculations. We can display the full object by directly accessing the corresponding list element, which in this case is:
\begin{tiny}
\begin{verbatim}
> condndndn[1,1,1];

<
           d                           d
          --- U(x0, x1, x2, x3)       --- phi(x0, x1, x2, x3)
          dx0                         dx0
        - --------------------- + 1/2 -----------------------
                    2                            4
                   c                            c
\end{verbatim}
\end{tiny}
Hence, we see that the full object is still in memory. Only the reduce command truncates the full object and thereby affects subsequent calculations, as we can see from the following command sequence:  
\begin{tiny}
\begin{verbatim}
> red_condndndn(4);condndndn[1,1,1];

<
                          d
                         --- U(x0, x1, x2, x3)
                         dx0
                       - ---------------------
                                   2
                                  c
\end{verbatim}
\end{tiny}
This also shows that no derivative replacements are performed during the reduction of full objects. The maximum order up to which objects get truncated is controlled via the global variable \texttt{CCmaxorder} (cf.\ table \ref{global_switches}). The user should make sure that the order specified in this variable is high enough to guarantee that all unwanted higher order terms are removed. If the user wants to display a single component of an object including the derivative replacements the command \texttt{disp\_component}, cf.\ table \ref{function_list_table_0}, should be used. For the connection component from the previous examples this command yields  
\begin{tiny}
\begin{verbatim}
> disp_component('condndndn[1,1,1]',4);

<
                                            d
                                            -- U
                                            dt
                      condndndn[1, 1, 1], - ----
                                              3
                                             c
\end{verbatim}
\end{tiny}
Finally, we note that the reduction commands should be used with great care, since they directly influence the calculation of subsequent objects if these depend on the object being reduced. 

\subsection{Saving and loading components}
Time and memory intensive calculations are supported by the package via an integrated save function, which is controlled by the global switch \texttt{CCoutputfs} (see table \ref{global_switches}). If this switch is set to the value $2$, all functions produce a file with reusable output. The calculation routines produce output files with names \texttt{<object/component name>.full.mpi}. Files with the names \texttt{<object/ component name>.red.mpi} are generated by the reduction routines. Output files are generated for every component of an indexed object and there are routines for loading the entire object as well as only single components. The sample worksheet contains a section in which the previously computed results are loaded from files. For all objects there exists a corresponding load routine \texttt{load\_<object name>(full/reduced switch)} which reads the complete object from files, the switch passed to the load command determines whether files with the suffix \texttt{.full.mpi} or \texttt{.red.mpi} are used (see table \ref{function_list_table_4} for a complete list). 

Components of objects can be loaded from the existing files by using the \texttt{load\_component} command. In contrast to the \texttt{load\_<object name>} command the \texttt{load\_component} command returns the specified component of an object but does {\it not} automatically assign it to the corresponding variable name. Here are some examples for objects with one and two indices and a scalar quantity: 
\begin{tiny}
\begin{verbatim}
> load_component('uu',[1],1); load_component('tupup',[1,2],2); load_component('thetascalar',[],2);

<                          "uu_1.full.mpi"
                           "tupup_1_2.red.mpi"
                           "thetascalar.red.mpi"
\end{verbatim}
\end{tiny}
As noted before the above commands do not automatically perform assignments to the corresponding variable names. This can be easily done by hand as follows
\begin{tiny}
\begin{verbatim}
> for aa from 1 to 4 do test_uu[aa]:=load_component('uu',[aa],2) od:
  
<                           "uu_1.red.mpi"
                            "uu_2.red.mpi"
                            "uu_3.red.mpi"
                            "uu_4.red.mpi"
\end{verbatim}
\end{tiny}
or by using the predefined load command, which loads all components of \texttt{uu} from the corresponding files and assigns them. We can verify the equivalence of the last command sequence and the load command by
\begin{tiny}
\begin{verbatim}
> load_uu(2);

<                           "uu_1.red.mpi"
                            "uu_2.red.mpi"
                            "uu_3.red.mpi"
                            "uu_4.red.mpi"

> seq(is(test_uu[aa]=uu[aa]),aa=1..4); 

<                       true, true, true, true
\end{verbatim}
\end{tiny}
More examples can be found in the sample worksheet. The complete list of predefined load commands is given in table \ref{function_list_table_4}. 

\subsection{Testrun}
After the installation it is advisable to perform a quick testrun with the package. This can easily be done by executing the worksheet supplied with the package. After performing the sample calculations (which should not take longer than a couple of minutes on a modern desktop machine\footnote{With Maple 8 it takes about 35 seconds on a Pentium M (1.7 GHz) system and approximately 320 MBytes of main memory.}) the worksheet automatically compares its results with some test data files which are supplied together with the package (as noted before the data for comparison is located in the directory \texttt{testrun\_procrustes<version number>}). All of the current features of the package are demonstrated in the sample worksheet. We note that the calculation in the sample worksheet reproduces many results of the classic work \cite{Chandrasekhar:1965:2} and is therefore a good example for the effectiveness of the package. 

\section{Conclusions and outlook}
\label{conclusions_sec}
We have introduced the main features of the Procrustes package and explained how it can be used to tackle computations within the post-Newtonian approximation of General Relativity. The package is versatile and should allow the user to easily make modifications. Possible future developments of the package strongly depend on the scientific problems in which the author will be involved. Planned future developments include: (i) An automated order control for all quantities entering a specific post-Newtonian calculation. This feature would further reduce the memory requirements of the program and would alleviate the user from making choices for the correct post-Newtonian orders of single quantities. (ii) Automated simplification schemes for the output. The goal is to introduce some extra routines which match certain patterns in the output and thereby reduce the need for manual calculations even further. (iii) Extension of the package to incorporate non-Riemannian theories of gravitation. The complexity of the geometric quantities in such theories renders virtually any hand calculation unfeasible. From a physical point of view this is one of the most pressing steps, since it would allow us to establish a systematic post-Newtonian framework for alternative gravitational theories.  

\section*{Acknowledgements}
The author is grateful to M.\ Pohl (Iowa State University) for his longterm support of this project. In addition the support by {\O}.\ Elgar{\o}y (University of Oslo) and the Research Council of Norway (project number 162830) during the later stages of this work is gratefully acknowledged. For the supply and support of different Maple versions the author would like to thank J.\ Veverka (Iowa State University), T.\ Leifsen (University of Oslo), and U.\ Fuskeland (University of Oslo). Finally, the author is grateful to two anonymous referees whose constructive comments helped to improve the manuscript. 

\appendix

\section{Directory of functions and names}
\label{functions_sec}

In this section we provide the names and a description of predefined switches \ref{global_switches}, a list of routines and their dependencies \ref{function_list_table_0}-\ref{function_list_table_4}, and a list of global variables \ref{global_variables} in table form.    

\begin{table}
\caption{List of global switches.}
\label{global_switches} 
\begin{center}
\begin{tabular}{p{2cm}p{0.01cm}p{0.01cm}p{10cm}}
\hline 
Name &&&Description\\
\hline 
\tiny \verb#CCmaxorder#&&&Order up to which expanded objects get truncated. The standard setting is \texttt{200}, meaning that the reduction of objects is performed starting from the order passed to the routine up to $c^{-200}$\\
\tiny \verb#CCoutputfs#&&&Switch which toggles the output to files. The standard setting is \texttt{2}, meaning that all routines will produce files with their output\\
\tiny \verb#CCprocrustes_dir#&&&Directory in which the routines of the package are stored, the standard setting is \texttt{"./procrustes<versionumber>/"}\\
\tiny \verb#CCtaylor#&&&Order variable used in Taylor approximations, the standard setting is \texttt{3}\\
\hline
\end{tabular}
\end{center}
\end{table}

\begin{table}
\caption{List of general predefined functions.}
\label{function_list_table_0} 
\begin{center}
\begin{tabular}{p{7cm}p{6cm}}
\hline 
Name&Description\\
\hline 
\tiny \verb#system_status()#&Outputs memory usage of the current session\\
\tiny \verb#fapprox(<expression>,<order>,<point>)#&Taylor approximation of the inverse square root up to the given order at a given point\\
\tiny \verb#gapprox(<expression>,<order>,<point>)#&Taylor approximation of the square root up to the given order at a given point\\
\tiny \verb#disp_component('<object name>',<display order>)#&Displays an object in memory up to the given order \\
\tiny \verb#load_component(<object name>,<index>,<full/reduced expression>)#&Loads object components from file, last switch used to toggle between full (``=1") / reduced (``=2") expressions\\
\hline 
\end{tabular}
\end{center}
\end{table}

\begin{table}
\caption{List of predefined functions which calculate an object and display it up to the given \texttt{<display order>}.}
\label{function_list_table_1} 
\begin{center}
\begin{tabular}{p{3.8cm}p{2cm}p{7cm}}
\hline 
Name  &Dependencies&Description\\
\hline 
\tiny \verb#calc_uu(<display order>)#&\tiny \verb#gdd, fapprox()#&Contravariant four velocity $u^\alpha=dx^\alpha/ds$\\
\tiny \verb#calc_ud(<display order>)#&\tiny \verb#gdd, uu#&Covariant four velocity $u_\alpha=g_{\alpha \beta} u^{\beta}$\\
\tiny \verb#calc_tupup(<display order>)#&\tiny \verb#uu, guu#&Contravariant perfect fluid energy-momentum tensor $T^{\alpha \beta}$\\
\tiny \verb#calc_tdndn(<display order>)#&\tiny \verb#ud, gdd#&Covariant perfect fluid energy-momentum tensor $T_{\alpha \beta }=\left( \rho c^{2}+\Pi \rho +p\right) u_{\alpha }u_{\beta}-pg_{\alpha \beta }$\\
\tiny \verb#calc_tupdn(<display order>)#&\tiny \verb#gdd, tupup#&Mixed perfect fluid energy-momentum tensor $T^{\alpha}{}_{\beta}$\\
\tiny \verb#calc_rhs(<display order>)#&\tiny \verb#gdd, tdndn, tupdn#&Right-hand side of the field equations ${\rm RHS}_{\alpha \beta }=-\frac{\kappa}{c^{4}}\left( T_{\alpha \beta }-\frac{1}{2}T^{\gamma }{}_{\gamma }g_{\alpha \beta }\right)$\\
\tiny \verb#calc_condndndn(<display order>)#&\tiny \verb#gdd#&Connection $\Gamma _{\gamma |\alpha \beta }=\frac{1}{2}\left( g_{\gamma \alpha ,\beta}+g_{\beta \gamma ,\alpha }\right.$ $\left. -g_{\alpha \beta ,\gamma }\right)$\\
\tiny \verb#calc_conupdndn(<display order>)#&\tiny \verb#guu, condndndn#&Connection $\Gamma^{\gamma}_{\alpha \beta } = g^{\gamma \delta} \Gamma _{\delta |\alpha \beta }$\\
\tiny \verb#calc_ctupup(<display order>)#&\tiny \verb#tupup, conupdndn#&Covariant derivative of the EM tensor $T^{\alpha \beta}{}_{;\gamma}$\\
\tiny \verb#calc_eom(<display order>)#&\tiny \verb#ctupup#&Equations of motion $T^{\alpha \beta}{}_{;\beta}=0$\\
\tiny \verb#calc_ricdndn(<display order>)#&\tiny \verb#guu, gdd, condndndn#&Covariant Ricci tensor $R_{\mu \nu} = \frac{1}{2}g^{\alpha \beta } ( g_{\mu \nu ,\alpha \beta } + g_{\alpha \beta ,\mu \nu } - g_{\alpha \nu ,\mu \beta } - g_{\mu \beta ,\alpha \nu } ) + g^{\alpha \beta }g^{\gamma \delta } ( \Gamma _{\delta |\mu \nu} \Gamma _{\gamma |\alpha \beta } - \Gamma _{\gamma |\mu \beta } \Gamma _{\delta |\alpha \nu })$\\
\tiny \verb#calc_ricupdn(<display order>)#&\tiny \verb#guu, ricdndn#&Mixed Ricci tensor $R^{\alpha}{}_{ \beta}$\\
\tiny \verb#calc_ricscalar(<display order>)#&\tiny \verb#ricupdn#&Mixed Ricci scalar $R=R^{\alpha}{}_{ \alpha}$\\
\tiny \verb#calc_riemupdndndn(<display order>)#&\tiny \verb#conupdndn#&Riemann curvature tensor $R^{\alpha}{}_{ \beta \mu \nu}=-\Gamma _{\beta \mu ,\nu }^{\alpha }+\Gamma_{\beta \nu ,\mu }^{\alpha }-\Gamma _{\beta \mu }^{\gamma }\Gamma _{\gamma \nu }^{\alpha }+\Gamma _{\beta \nu }^{\gamma }\Gamma _{\gamma \mu }^{\alpha }$\\
\tiny \verb#calc_riemdndndndn(<display order>)#&\tiny \verb#gdd, riemupdndn#&Riemann curvature tensor $R_{\alpha \beta \mu \nu}=g_{\alpha \gamma} R^{\gamma}{}_{ \beta \mu \nu}$\\
\tiny \verb#calc_weyldndndndn(<display order>)#&\tiny \verb#gdd, ricdndn,# \tiny \verb#ricscalar,# \tiny \verb#riemdndndndn#&Weyl curvature tensor $C_{\alpha \beta \mu \nu}=R_{\alpha \beta \mu \nu }-\frac{1}{2} (g_{\alpha \nu }B_{\beta \mu }+g_{\beta \mu }B_{\alpha \nu }-g_{\alpha \mu}B_{\beta \nu }-g_{\beta \nu }B_{\alpha \mu }) -\frac{1}{12}R (g_{\alpha \nu }g_{\beta \mu }-g_{\alpha \mu }g_{\beta \nu })$ with $B_{\alpha \beta }=R_{\alpha \beta }-\frac{1}{4}g_{\alpha \beta }R$\\
\tiny \verb#calc_detguu(<display order>)#&\tiny \verb#guu#&Metric determinant $det(g^{\alpha \beta})$\\
\tiny \verb#calc_detgdd(<display order>)#&\tiny \verb#gdd#&Metric determinant $det(g_{\alpha \beta})$\\
\hline 
\end{tabular}
\end{center}
\end{table}

\begin{table}
\caption{List of predefined functions which calculate an object and display it up to the given \texttt{<display order>}.}
\label{function_list_table_2} 
\begin{center}
\begin{tabular}{p{3.8cm}p{2cm}p{7cm}}
\hline
Name&Dependencies&Description\\
\hline  
 \tiny \verb#calc_pemtupup(<display order>)#&\tiny \verb#guu, conupdndn#&Landau-Lifshitz energy-momentum pseudotensor 
$t^{\mu \nu } = \frac{c^{4}}{2 \kappa} [ ( 2\Gamma _{\alpha \beta}^{\delta }\Gamma _{\delta \kappa }^{\kappa }-\Gamma _{\alpha \kappa}^{\delta }\Gamma _{\beta \delta }^{\kappa }-\Gamma _{\alpha \delta}^{\delta }\Gamma _{\beta \kappa }^{\kappa }) ( g^{\mu \alpha}g^{\nu \beta }-g^{\mu \nu }g^{\alpha \beta }) +g^{\mu \alpha }g^{\beta \delta }( \Gamma _{\alpha \kappa }^{\nu}\Gamma _{\beta \delta }^{\kappa }+\Gamma _{\beta \delta }^{\nu }\Gamma_{\alpha \kappa }^{\kappa }-\Gamma _{\delta \kappa }^{\nu }\Gamma _{\alpha\beta }^{\kappa }-\Gamma _{\alpha \beta }^{\nu }\Gamma _{\delta \kappa}^{\kappa })+g^{\nu \alpha }g^{\beta \delta }( \Gamma _{\alpha \kappa }^{\mu}\Gamma _{\beta \delta }^{\kappa }+\Gamma _{\beta \delta }^{\mu }\Gamma_{\alpha \kappa }^{\kappa }-\Gamma _{\delta \kappa }^{\mu }\Gamma _{\alpha\beta }^{\kappa }-\Gamma _{\alpha \beta }^{\mu }\Gamma _{\delta \kappa}^{\kappa })+g^{\alpha \beta }g^{\delta \kappa }( \Gamma _{\alpha \delta}^{\mu }\Gamma _{\beta \kappa }^{\nu }-\Gamma _{\alpha \beta }^{\mu }\Gamma_{\delta \kappa }^{\nu }) ]$\\
\tiny \verb#calc_thetaupup(<display order>)#&\tiny \verb#detgdd, tupup,# \tiny \verb#pemtupup#&Landau-Lifshitz energy-momentum complex  $\Theta ^{\alpha \beta }=-\det ( g_{\alpha \beta }) [ T^{\alpha \beta }+t^{\alpha \beta }]$\\
\tiny \verb#calc_thetaupuppdn(<display order>)#&\tiny \verb#thetaupup#&Partial derivative of the LL EM complex $\Theta^{\alpha \beta}{}_{,\beta}$\\
\tiny \verb#calc_harm_gc(<display order>)#&\tiny \verb#guu, detgdd,# \tiny \verb#gapprox()#&Harmonic gauge condition $(\sqrt{-g}g^{\alpha \beta})_{,\beta}$\\
\tiny \verb#calc_harm_gc_alt(<display order>)#&\tiny \verb#guu, conupdndn #&Harmonic gauge condition $g^{\alpha \beta} \Gamma^\gamma_{\alpha \beta}$ (alternative form)\\
\tiny \verb#calc_pn_gc(<display order>)#&\tiny \verb#gdd#&Standard post-Newtonian gauge condition ($\delta^{ab} ( g_{0a,b}-\frac{1}{2}g_{ab,0} )=0$, $\delta^{bc} g_{ab,c}-\frac{1}{2}(\delta^{bc} g_{bc}-g_{00})_{,a}=0$)\\
\tiny \verb#calc_cuu(<display order>)#&\tiny \verb#uu, conupdndn#&Covariant derivative of the four velocity $u^\alpha{}_{;\beta}$\\
\tiny \verb#calc_cud(<display order>)#&\tiny \verb#ud, conupdndn#&Covariant derivative of the four velocity $u_{\alpha;\beta}$\\
\tiny \verb#calc_thetascalar(<display order>)#&\tiny \verb#cuu#&Expansion scalar $\theta=u^\alpha{}_{;\alpha}$\\
\tiny \verb#calc_acd(<display order>)#&\tiny \verb#uu, cud#&Acceleration $a_\alpha=u_{\alpha;\beta}u^{\beta}$\\
\tiny \verb#calc_omegadd(<display order>)#&\tiny \verb#ud, cud, acd#&Rotation $\omega_{\alpha \beta}=u_{[\alpha;\beta]}+u_{[\alpha}a_{\beta]}$\\
\tiny \verb#calc_thetadd(<display order>)#&\tiny \verb#ud, cud, acd#&Symmetric part of the derivative projection $\theta_{\alpha \beta}=u_{(\alpha;\beta)}-u_{(\alpha}a_{\beta)}$\\
\tiny \verb#calc_sigmadd(<display order>)#&\tiny \verb#gdd, ud,# \tiny \verb#thetadd,# \tiny \verb#thetascalar#&Shear $\sigma_{\alpha \beta}=\theta_{\alpha \beta} - \frac{1}{3} \theta (g_{\alpha \beta} - u_\alpha u_\beta)$\\
\tiny \verb#check_velnorm(<display order>)#&\tiny \verb#uu, ud#&Checks fulfillment of the velocity normalization\\
\tiny \verb#check_metricid(<display order>)#&\tiny \verb#guu, gdd#&Checks fulfillment of the metric identity\\
\tiny \verb#check_decomposition(<display order>)#&\tiny \verb#omegadd, sigmadd,# \tiny \verb#thetascalar,# \verb#gdd,# \tiny \verb#ud, acd, cud#&Checks whether the decomposition $u_{\alpha;\beta}=\omega_{\alpha \beta}+\sigma_{\alpha \beta}+\frac{1}{3}(g_{\alpha \beta} - u_\alpha u_\beta) \theta +a_\alpha u_\beta$ is consistent\\
\hline 
\end{tabular}
\end{center}
\end{table}

\begin{table}
\caption{List of predefined functions which truncate an object at the given \texttt{<truncation order>}.}
\label{function_list_table_3} 
\begin{center}
\begin{tabular}{p{4cm}p{2cm}p{7cm}}
\hline 
Name&Dependencies&Object\\
\hline
\tiny \verb#red_uu(<truncation order>)#&\tiny \verb#uu#&Four velocity $u^\alpha$\\
\tiny \verb#red_ud(<truncation order>)#&\tiny \verb#ud#&Four velocity $u_\alpha$\\
\tiny \verb#red_tupup(<truncation order>)#&\tiny \verb#tupup#&EM tensor $T^{\alpha \beta}$\\
\tiny \verb#red_tdndn(<truncation order>)#&\tiny \verb#tdndn#&EM tensor $T_{\alpha \beta}$\\
\tiny \verb#red_tupdn(<truncation order>)#&\tiny \verb#tupdn#&EM tensor $T^{\alpha}{}_{\beta}$\\
\tiny \verb#red_rhs(<truncation order>)#&\tiny \verb#rhsfeqsdndn#&Right-hand side of FEQs ${\rm RHS}_{\alpha \beta }$\\
\tiny \verb#red_condndndn(<truncation order>)#&\tiny \verb#condndndn#&Connection $\Gamma _{\gamma |\alpha \beta }$\\
\tiny \verb#red_conupdndn(<truncation order>)#&\tiny \verb#conupdndn#&Connection $\Gamma^{\gamma}_{\alpha \beta }$\\
\tiny \verb#red_ctupup(<truncation order>)#&\tiny \verb#ctupup#&Covariant deriv.\ of the EM tensor $T^{\alpha \beta}{}_{;\gamma}$\\
\tiny \verb#red_eom(<truncation order>)#&\tiny \verb#eom#&EOMs $T^{\alpha \beta}{}_{;\beta}=0$\\
\tiny \verb#red_ricdndn(<truncation order>)#&\tiny \verb#ricdndn#&Covariant Ricci tensor $R_{\alpha \beta }$\\
\tiny \verb#red_ricupdn(<truncation order>)#&\tiny \verb#ricupdn#&Mixed Ricci tensor $R^{\alpha}{}_{ \beta}$\\
\tiny \verb#red_ricscalar(<truncation order>)#&\tiny \verb#ricscalar#&Ricci scalar $R^{\alpha}{}_{ \alpha}$\\
\tiny \verb#red_riemupdndndn(<truncation order>)#&\tiny \verb#riemupdndn#&Riemann curvature tensor $R^{\alpha}{}_{ \beta \mu \nu}$\\
\tiny \verb#red_riemdndndndn(<truncation order>)#&\tiny \verb#riemdndndn#&Riemann curvature tensor $R_{\alpha \beta \mu \nu}$\\ 
\tiny \verb#red_weyldndndndn(<truncation order>)#&\tiny \verb#weyldndndndn#&Weyl curvature tensor $C_{\alpha \beta \mu \nu}$\\
\tiny \verb#red_detguu(<truncation order>)#&\tiny \verb#detguu#&Metric determinant $det(g^{\alpha \beta})$\\
\tiny \verb#red_detgdd(<truncation order>)#&\tiny \verb#detgdd#&Metric determinant $det(g_{\alpha \beta})$\\
\tiny \verb#red_pemtupup(<truncation order>)#&\tiny \verb#pemtupup#&LL EMPT $t^{\alpha \beta}$\\
\tiny \verb#red_thetaupup(<truncation order>)#&\tiny \verb#thetaupup#&LL EM complex $\Theta^{\alpha \beta}$\\
\tiny \verb#red_thetaupuppdn(<truncation order>)#&\tiny \verb#thetaupuppdn#&Partial deriv.\ of the LL EM complex $\Theta^{\alpha \beta}{}_{,\beta}$ \\
\tiny \verb#red_cuu(<truncation order>)#&\tiny \verb#cuu#&Covariant derivative $u^\alpha{}_{;\beta}$\\
\tiny \verb#red_cud(<truncation order>)#&\tiny \verb#cud#&Covariant derivative $u_{\alpha;\beta}$\\
\tiny \verb#red_thetascalar(<truncation order>)#&\tiny \verb#thetascalar#&Expansion scalar $\theta$\\
\tiny \verb#red_acd(<truncation order>)#&\tiny \verb#acd#&Acceleration $a_\alpha$\\
\tiny \verb#red_omegadd(<truncation order>)#&\tiny \verb#omegadd#&Rotation $\omega_{\alpha \beta}$\\
\tiny \verb#red_thetadd(<truncation order>)#&\tiny \verb#thetadd#&Symm.\ part of the deriv.\ projection $\theta_{\alpha \beta}$\\
\tiny \verb#red_sigmadd(<truncation order>)#&\tiny \verb#sigmadd#&Shear $\sigma_{\alpha \beta}$\\
\hline   
\end{tabular}
\end{center}
\end{table}

\begin{table}
\caption{List of predefined functions which load objects from files. The parameter \texttt{<full/reduced expression>} passed to the function determines whether the full or reduced object is loaded (\texttt{1 = <object name>.full.mpi, 2 = <object name>.red.mpi}). All functions depend on the \texttt{load\_component()} routine.}
\label{function_list_table_4} 
\begin{center}
\begin{tabular}{p{5cm}p{8cm}}
\hline 
Name&Object\\
\hline
\tiny \verb#load_uu(<full/reduced expression>)#&Four velocity $u^\alpha$\\
\tiny \verb#load_ud(<full/reduced expression>)#&Four velocity $u_\alpha$\\
\tiny \verb#load_tupup(<full/reduced expression>)#&EM tensor $T^{\alpha \beta}$\\
\tiny \verb#load_tdndn(<full/reduced expression>)#&EM tensor $T_{\alpha \beta}$\\
\tiny \verb#load_tupdn(<full/reduced expression>)#&EM tensor $T^{\alpha}{}_{\beta}$\\
\tiny \verb#load_rhs(<full/reduced expression>)#&Right-hand side of FEQs ${\rm RHS}_{\alpha \beta }$\\
\tiny \verb#load_condndndn(<full/reduced expression>)#&Connection $\Gamma _{\gamma |\alpha \beta }$\\
\tiny \verb#load_conupdndn(<full/reduced expression>)#&Connection $\Gamma^{\gamma}_{\alpha \beta }$\\
\tiny \verb#load_ctupup(<full/reduced expression>)#&Covariant derivative of the EM tensor $T^{\alpha \beta}{}_{;\gamma}$\\
\tiny \verb#load_eom(<full/reduced expression>)#&EOMs $T^{\alpha \beta}{}_{;\beta}=0$\\
\tiny \verb#load_ricdndn(<full/reduced expression>)#&Covariant Ricci tensor $R_{\alpha \beta }$\\
\tiny \verb#load_ricupdn(<full/reduced expression>)#&Mixed Ricci tensor $R^{\alpha}{}_{ \beta}$\\
\tiny \verb#load_ricscalar(<full/reduced expression>)#&Ricci scalar $R^{\alpha}{}_{ \beta}$\\
\tiny \verb#load_riemupdndndn(<full/reduced expression>)#&Riemann curvature tensor $R^{\alpha}{}_{ \beta \mu \nu}$\\
\tiny \verb#load_riemdndndndn(<full/reduced expression>)#&Riemann curvature tensor $R_{\alpha \beta \mu \nu}$\\
\tiny \verb#load_weyldndndndn(<full/reduced expression>)#&Weyl curvature tensor $C_{\alpha \beta \mu \nu}$\\
\tiny \verb#load_detguu(<full/reduced expression>)#&Metric determinant $det(g^{\alpha \beta})$\\
\tiny \verb#load_detgdd(<full/reduced expression>)#&Metric determinant $det(g_{\alpha \beta})$\\
\tiny \verb#load_pemtupup(<full/reduced expression>)#&LL EMPT $t^{\alpha \beta}$\\
\tiny \verb#load_thetaupup(<full/reduced expression>)#&LL EM complex $\Theta^{\alpha \beta}$\\
\tiny \verb#load_thetaupuppdn(<full/reduced expression>)#&Partial derivative of the LL EM complex $\Theta^{\alpha \beta}{}_{,\beta}$\\
\tiny \verb#load_cuu(<full/reduced expression>)#&Covariant velocity derivative $u^\alpha{}_{;\beta}$\\
\tiny \verb#load_cud(<full/reduced expression>)#&Covariant velocity derivative $u_{\alpha;\beta}$\\
\tiny \verb#load_thetascalar(<full/reduced expression>)#&Expansion scalar $\theta$\\
\tiny \verb#load_acd(<full/reduced expression>)#&Acceleration $a_\alpha$\\
\tiny \verb#load_omegadd(<full/reduced expression>)#&Rotation $\omega_{\alpha \beta}$\\
\tiny \verb#load_thetadd(<full/reduced expression>)#&Symmetric part of the derivative projection $\theta_{\alpha \beta}$\\
\tiny \verb#load_sigmadd(<full/reduced expression>)#&Shear $\sigma_{\alpha \beta}$\\
\hline
\end{tabular}
\end{center}
\end{table}

\begin{table}
\caption{List of global variables}
\label{global_variables} 
\begin{center}
\begin{tabular}{p{2cm}p{0.01cm}p{0.01cm}p{10cm}}
\hline 
Name &&&Description\\
\hline
\tiny \verb#acd#&&&Kinematic acceleration $a_\alpha$\\
\tiny \verb#condndndn, conupdndn#&&&Connection $\Gamma_{\gamma | \alpha \beta}$, $\Gamma^{\alpha}_{\beta \gamma}$\\
\tiny \verb#ctupup#&&&Covariant derivative of the EM tensor $T^{\alpha \beta}{}_{;\gamma}$\\
\tiny \verb#cuu, cud#&&&Covariant derivative of the four velocity $u^\alpha{}_{;\beta}$, $u_{\alpha;\beta}$\\
\tiny \verb#detgdd, detguu#&&&Metric determinant $det(g_{\alpha \beta})$, $det(g^{\alpha \beta})$\\
\tiny \verb#displist#&&&Substitution list for replacements in output\\
\tiny \verb#eom#&&&Equations of motion $T^{\alpha \beta}{}_{;\beta}$\\
\tiny \verb#gdd, guu#&&&Covariant $g_{\alpha \beta}$ / contravariant $g^{\alpha \beta}$ metric\\
\tiny \verb#harm_gc, harm_gc_alt#&&&Harmonic gauge condition\\
\tiny \verb#omegadd#&&&Kinematic rotation $\omega_{\alpha \beta}$\\
\tiny \verb#pemtupup#&&&Landau-Lifshitz EM pseudotensor $t^{\alpha \beta}$\\
\tiny \verb#pn_gc#&&&Standard post-Newtonian gauge condition\\
\tiny \verb#ricdndn, ricupdn#&&&Ricci tensor $R_{\alpha \beta}$, $R^{\alpha}{}_{\beta}$\\
\tiny \verb#ricscalar#&&&Ricci scalar $R$\\
\tiny \verb#riemupdndndn, riemdndndndn#&&&Riemann curvature tensor $R^{\alpha}{}_{\beta \gamma \delta}$, $R_{\alpha \beta \gamma \delta}$\\
\tiny \verb#rhsfeqsdndn#&&&Right hand side of the field equations $-\frac{\kappa}{c^{4}}\left( T_{\alpha \beta }-\frac{1}{2}T^{\gamma }{}_{\gamma }g_{\alpha \beta }\right)$\\
\tiny \verb#routine_list#&&&List containing the names of routines to be loaded from routine directory\\
\tiny \verb#sigmadd#&&&Kinematic shear $\sigma_{\alpha \beta}$\\
\tiny \verb#sublist#&&&Substitution list for derivative replacements\\
\tiny \verb#tdndn, tupup, tupdn#&&&Energy-momentum tensor $T_{\alpha \beta}$, $T^{\alpha \beta}$, $T^{\alpha}{}_{\beta}$\\
\tiny \verb#thetadd#&&&Kinematic quantity $\theta_{\alpha \beta}$, symmetric part of the velocity derivative projection\\
\tiny \verb#thetaupup, thetaupuppdn#&&&Landau-Lifshitz EM complex $\Theta^{\alpha \beta}$ and its partial derivative $\Theta^{\alpha \beta}{}_{,\beta}$\\
\tiny \verb#thetascalar#&&&Kinematic expansion $\theta$\\
\tiny \verb#ud, uu#&&&Covariant $u_{\alpha}$ / contravariant $u^{\alpha}$ four velocity\\
\tiny \verb#velnorm#&&&Velocity product $u^\alpha u_\alpha$\\
\tiny \verb#weyldndndndn#&&&Weyl curvature tensor $C_{\alpha \beta \gamma \delta}$\\
\tiny \verb#xx#&&&List with coordinates\\
\hline
\end{tabular}
\end{center}
\end{table}

\bibliographystyle{plain}
\bibliography{./procrustes_bibliography}

\end{document}